\documentclass{sig}

\usepackage{amssymb}
\usepackage{url}

\newcommand{\hidecomment}[1]{}
\newcommand{\myparagraph}[1]{\noindent \textbf{#1}}

\newenvironment{definition}
{\stepcounter{definition} \textbf{Definition \arabic{definition}}:}
{ \hspace{\stretch{1}}\rule{1ex}{1ex}}

\newcounter{definition}

{\begin{list}{$\bullet$}{%
        \setlength{\labelsep}{2pt}\setlength{\leftmargin}{0pt}%
        \setlength{\labelwidth}{0pt}%
        \setlength{\listparindent}{0pt}%
        \setlength{\itemsep}{.1\itemsep}}}
{\end{list}}

\begin{document}
\title{Pooling Hybrid Representations for \\ Web Structured Data Annotation}

\numberofauthors{3}
\author{
\alignauthor Luciano Barbosa \\
       \affaddr{Universidade Federal de Pernambuco}\\
       \email{luciano@cin.ufpe.br}
\and
\alignauthor Breno W. Carvalho \\
       \affaddr{IBM Research - Brazil}\\
       \email{brenow@br.ibm.com}
\and
\alignauthor Bianca Zadrozny \\
       \affaddr{IBM Research - Brazil}\\
       \email{biancaz@br.ibm.com}
}       

\maketitle

\begin{abstract}
Automatically identifying data types of web structured data is
a key step in the process of web data integration. Web structured data is usually associated with entities or objects in a particular domain. In this paper, we aim to map attributes of an entity in a given domain to pre-specified classes of attributes in the same domain based on their values. To perform this task, we propose a hybrid deep learning network that relies on the format of the attributes' values. It does so without any pre-processing or using pre-defined hand-crafted features.  
The hybrid network combines sequence-based neural networks, namely convolutional neural networks (CNN) and recurrent neural networks (RNN), to learn the sequence structure of attributes' values. The CNN captures short-distance dependencies in these sequences through a sliding window approach, and the RNN captures long-distance dependencies by storing information of previous characters. These networks create different vector representations of the input sequence which are combined using a pooling layer. This layer applies a specific operation on these vectors in order to capture their most useful patterns for the task. Finally, on top of the pooling layer, a softmax function predicts the label of a given attribute value.
We evaluate our strategy in four different web domains. The results show that the pooling network outperforms previous approaches, which use some kind of input pre-processing, in all domains.
\end{abstract}

\section{Introduction}
\begin{figure}[t]
	\begin{center}
	\includegraphics[width=.9\linewidth]{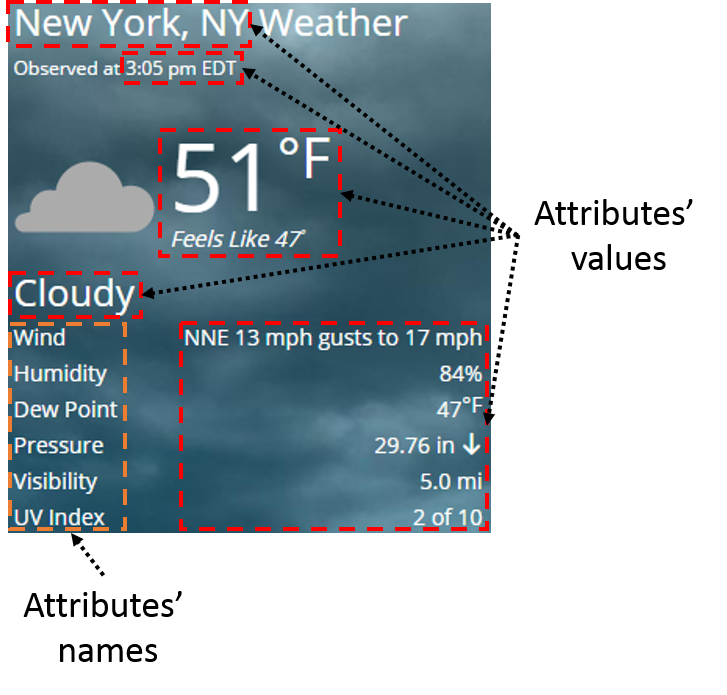}
	\end{center}
	\caption{Web page of the weather.com website showing attributes' names and values of a weather-related entity.}
	\label{fig:example}
\end{figure}

The Web is a rich source of structured data. 
Examples of Web structured data are Wikipedia infoboxes, the content of online databases~\cite{cafarella@sigrecord2008} and specification of products~\cite{crestan@wsm2011}. Typically, this data represents attributes (and their respective values) of entities or objects. Figure~\ref{fig:example} shows an example of a weather-related entity with its attributes' names and values highlighted.

Web structured data is currently used in a wide range of applications. Search engines use structured data to augment their search results with structured information as a response to a queried entity (e.g., movies and actors). Entities and relations extracted from Web structured data are also used to build knowledge bases~\cite{cafarella@cacm2011}. Moreover, the extracted information itself, not only services created from it, has been made available on data-as-a-service market sites\footnote{Examples of data market sites: https://www.factual.com/ and http://www.xignite.com/}.

To take advantage of such useful content, many approaches have been proposed to collect~\cite{qiu@vldb2015,meusel@cikm2014}, extract~\cite{gupta@vldb2009,cafarella@vldb2008} and integrate~\cite{madhavan@cidr2007} Web structured data.
A key step in the process of web data integration is to automatically label attributes of entities in online sources. Consider, for instance, the example of the weather entity in Figure~\ref{fig:example}. Although some of its attributes have explicit labels such as ``wind'', ``humidity'' and ``dew point'', others have only their values presented as, for instance, the value ``Cloudy'' for the attribute weather condition, or the value ``3:05 pm EDT'' for time. Attribute annotation is essential for better understanding the semantics of a given entity, and to map its attributes into a global schema that enables integration of entities from different sources.

In this paper, we focus on the problem of attribute annotation based only on the values of the attributes. The main challenge in performing this task is the variability of content and format of attributes' values. For instance, in the car search domain there are textual attributes (e.g. the interior color and transmission of the car), and numeric ones (e.g. mpg and number of doors). Even for values of the same attribute, there is a wide variability in how they are represented. As an example, the value of the attribute ``miles per gallon'' in the website $carsforsale.com$ is represented as ``21/32'', whereas in the site $carfax.com$ the same attribute is shown as `21 city / 32 hwy EPA Fuel Economy Guide''.

Previous approaches~\cite{cortez2010ondux,goel2012exploiting} for the problem of attribute annotation have used probabilistic frameworks that relied on hand-crafted features~\cite{goel2012exploiting}, or a modification of \emph{tfidf}~\cite{salton1986introduction} for textual values and statistical tests for numerical values~\cite{cortez2010ondux}. 
We aim to tackle this problem, using deep learning (DL) approaches. DL techniques have obtained state-of-the-art results in a great variety of text-related problems as, for instance, named entity recognition tagging~\cite{collobert@2011}, sentence 
classification~\cite{kim1@emnlp2014} and machine translation~\cite{sutskever@nips2014}. As opposed to previous approaches for attribute annotation, our DL network performs this task using as input only the raw text of the attributes' values: no pre-processing is required on the input nor pre-defined hand-crafted features are used. This makes the applicability of our strategy much easier, handling transparently numerical and textual values.

Hybrid deep learning networks have been proposed for computer vision~\cite{lin@iccv2015}, video recognition~\cite{simonyan@nips2014} and sentence similarity~\cite{santos@acl2015} problems. Similar to what was done for these tasks, we propose to combine different types of DL networks for the task of attribute annotation. More specifically, our network is a combination of two sequence-based neural networks, namely, a convolutional neural network~\cite{lecun@ieee1998} (CNN) and a recurrent neural network~\cite{cruse2006neural} (RNN). The advantage of using such a hybrid representation is that it allows capturing information at different granularity levels (short-distance and long-distance) from the input.

\begin{figure}[t]
	\begin{center}
	\includegraphics[width=.6\linewidth]{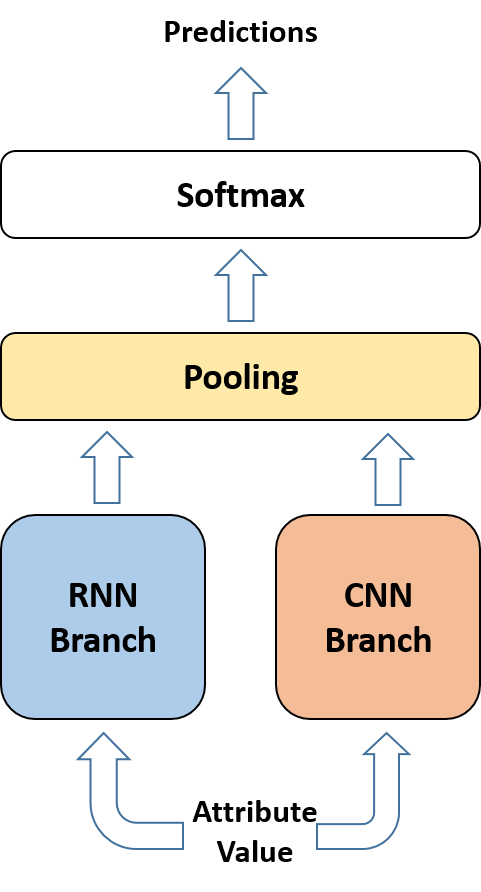}
	\end{center}
	\caption{Overview of our pooling-based hybrid network. The RNN and CNN branches create different representations of the same attribute value, a pooling layer then combines them using a pre-specified operation, and finally a softmax function predicts the label of the attribute value.}
	\label{fig:overview}
\end{figure}

Figure~\ref{fig:overview} gives an overview of our proposed network. 
Both a CNN and an RNN independently project the input into 
different feature spaces. The CNN branch captures short-distance dependencies in these sequences through a sliding window approach, whereas  the RNN branch captures long-distance dependencies by storing information about previous states (or characters in our context). 
The CNN version used here is the convolutional max-over-time network~\cite{collobert@2011}, which is more suitable for text-related tasks. The RNN version is the long short-term memory (LSTM)~\cite{hochreiter@nn1997} that deals with gradient issues present in the original RNN. 
These branches create different vector representations of the input sequences, which are combined using a pooling layer. The pooling layer applies a specific operation (e.g. max, sum or outer product) over these vectors in order to capture their most useful patterns. We have implemented and tested a set of pooling operations to be applied in this layer. Finally, on top of the pooling layer, a softmax function predicts the label of the attribute value. 

We have performed an extensive experimental evaluation using data from four different web domains: car search, weather sites, phone directories and bibliography citations. We have compared our proposed pooling network against state-of-the-art approaches from the literature and variations of the deep learning architecture. The results show that the pooling network outperformed the previous approaches in all domains. 

In summary, the main contributions of this paper are:
\begin{itemize}
\item Proposing the use of deep learning techniques for the task of attribute annotation, eliminating the need for  input pre-processing or hand-crafted features.
\item Introducing a novel pooling strategy that combines two sequence-based deep learning networks to encode different representations of character sequences, achieving state-of-the art results for the attribute annotation task.
\end{itemize}

The rest of the paper is organized as follows. Section~\ref{sec:problem} presents basic definitions and states our problem. Section~\ref{sec:network} introduces in details our hybrid network for attribute annotation. In Section~\ref{sec:evaluation}, we evaluate our approach and compare it with baselines. Finally, in Section~\ref{sec:related} we discuss related work and conclude in Section~\ref{sec:conc}.

\section{Problem Setting}\label{sec:problem}

We define an entity as follows:

\begin{definition} [Entity]
An entity $E$ is an object composed of a set of attribute-value pairs $\{<a_i,v_i>,...,<a_n,v_n>\}$.
\end{definition}

\begin{definition} [Domain] A domain $D$ is a set of entities that share common attributes.
\end{definition}

To give a concrete example of these definitions, a car advertisement, specified by its attributes, such as price, mileage and number of doors, is an example of an entity. The enitities of car advertisement belong to the same domain (car domain) since they share similar attributes and values. 

\begin{definition} [Domain Catalog] The Domain Catalog $DC$ specifies the attributes (or schema) of $D$ and values associated with each of its attributes.
\end{definition}

\begin{definition} [Problem Definition] Given a Domain Catalog $DC$ and an entity $E \in D$, we aim to map the attributes of $E$ to attributes in $DC$ based only on their values. Attributes are mapped if they have the same meaning in $D$.
\end{definition}

Given our supervised learning setup, the Domain Catalog provides the training data, in which each attribute is considered to be a class for a  classification task, and its values are the instances of that class used for training. At test time, the supervised learning model predicts the attributes of a given entity $E$ based on $DC$. It is important to point out we are not doing schema matching between columns of tables but between columns (or classes) in $DC$ and attributes of a single instance of $E$ in $D$ extracted from an online source.

\begin{figure}[t]
	\begin{center}
	\includegraphics[width=.9\linewidth]{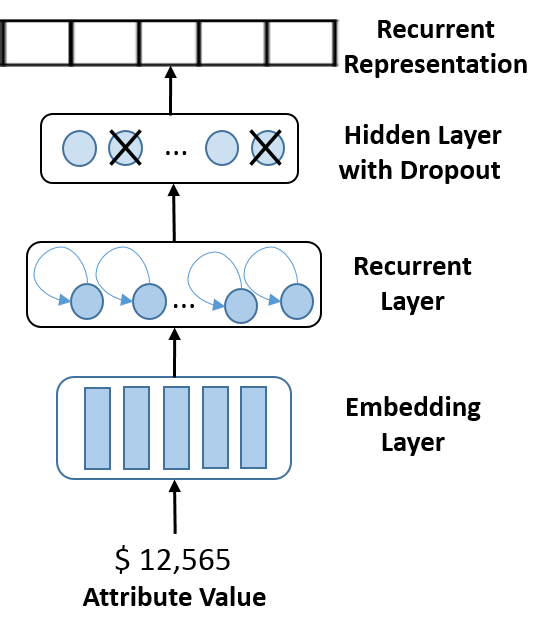}
	\end{center}
	\caption{Overview of the RNN Branch. First the embedding layer maps the attribute value to a lower dimensional space, then a recurrent layer followed by a hidden layer with dropout produces the recurrent representation of the given attribute value based on its sequence structure.}
	\label{fig:rnn_stream}
\end{figure}

\begin{figure}[t]
	\begin{center}
	\includegraphics[width=.9\linewidth]{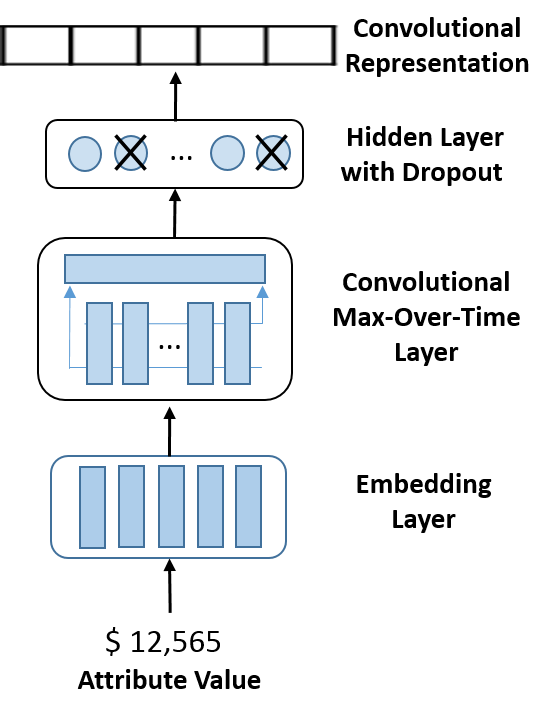}
	\end{center}
	\caption{Overview of the CNN Branch. First the embedding layer maps the attribute value to a lower dimensional space, then a convolutional max-over-time layer followed by a hidden layer with dropout produces the convolutional representation of the given attribute value based on its sequence structure.}
	\label{fig:cnn_stream}
\end{figure}

\section{Pooling Networks}\label{sec:network}
In this section, we present our solution for annotating web structured data.
Given the raw text of the attribute value, the network produces two different representations of the input using the recurrent and the convolutional branches (Figures~\ref{fig:rnn_stream} and~\ref{fig:cnn_stream}). No normalization or cleaning is performed on the input. These representations are then combined in a pooling layer that sub-samples them using a pre-defined operation (e.g. max, sum, average or multiplication). At the top of the network, a softmax layer applies a softmax function to the output vector of the pooling layer. The output of the softmax layer is a $k$-size vector containing the class-membership probabilities of the given input, where $k$ is the number of classes of attributes. In the remaining of this section, we present further details of our approach. 

\subsection{Embedding Layer}
The first layer of both branches is the embedding layer~\cite{collobert@2011}. Originally, embeddings have been applied for dimensionality reduction in the context of words: a word in a given large vocabulary is mapped to a vector in a lower dimensional space (e.g. 100 or 200 dimensions). In our context, instead of representing words, we use embeddings to represent the characters that compose the attribute value, i.e., each character is mapped to a vector in a lower dimensional space. More formally:
\begin{equation}
{EMB}_W(i) = W_i
\end{equation}
where $W \in \mathbb{R}^{d\times|D|}$ is the matrix that contains the embeddings of vector size $d$ of all characters in the dictionary $D$ of size $|D|$, and $W_i \in \mathbb{R}^d$ is the embedding vector of character $i \in D$. We add to $D$, a specific representation of the beginning of the sequence of characters and another representing the end. The weights of $W$ are randomly initialized and learned during backpropagation. The vector dimension $d$ is a user-defined hyper-parameter. 

\subsection{Convolutional Max-Over-Time Layer}
\begin{figure}[t]
	\begin{center}
	\includegraphics[width=.9\linewidth]{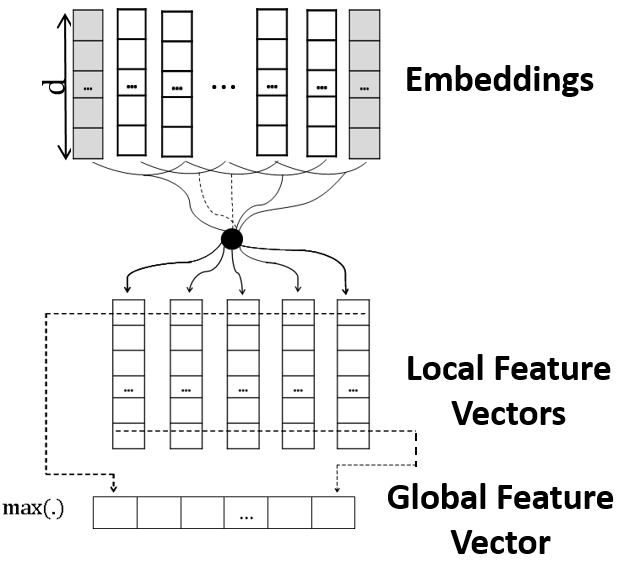}
	\end{center}
	\caption{Overview of the convolutional max-over-time layer. From the vector embeddings, the convolutional operation captures the local features using a sliding window, and then a max operator generates the global feature vector.}
	\label{fig:cnn}
\end{figure}

The convolutional max-over-time layer has been applied in different text-related tasks~\cite{collobert@2011,kim1@emnlp2014}. It captures local properties within a sequence of characters (convolution) and then applies an element-wise max operation producing a fixed-size vector (max pooling). Figure~\ref{fig:cnn} presents an overview of the convolutional max-over-time layer.

Given a sequence of vector embeddings $e=\{W_{c_1}, W_{c_2},...,$ $W_{c_n}\}$, which is the result of the mapping ${EMB}_W$ of a sequence of characters $s=\{c_1,c_2,...,c_n\}$, the convolution operation applies a filter $F \in  \mathbb{R}^{k \times d}$ to all sequences of $k$ characters (window size) in $e$, producing a sequence of $d$-size vectors $\{v_1,v_2,...,v_{n-(k-1)}\}$. More specifically, given a $k$-character sequence of embeddings $W_{c_{i:i+k-1}}$, the output of the convolution operation (d-vector $v_i$) over this sequence is the dot product of $F$ and $W_{c_{i:i+k-1}}$ plus a bias vector $b$:
\begin{equation}
 v_i = F \cdot W_{c_{i:i+k-1}} + b
\end{equation}

The resulting vectors of the convolution capture local properties of the $k$-character sequences. To combine these vectors in a global feature vector, we use an element-wise max operation. Each dimension of the global feature vector contains the maximum value for that dimension across all $v_i$ vectors. Another benefit of applying a max operation is that it results in a fixed-size vector even if the original character sequences have different sizes. The fixed-size output from the CNN layer can then be applied to regular neural network hidden layers or a softmax layer, which require fixed-size inputs. The window size $k$ is a hyper-parameter defined by the user.

\subsection{Long Short-Term Memory Layer}
Recurrent units allow the network to store information about previous states~\cite{cruse2006neural}. This is particularly helpful for sequence classification since sequence structure is preserved in the unit. 

\begin{figure}[t]
	\begin{center}
	\includegraphics[width=1\linewidth]{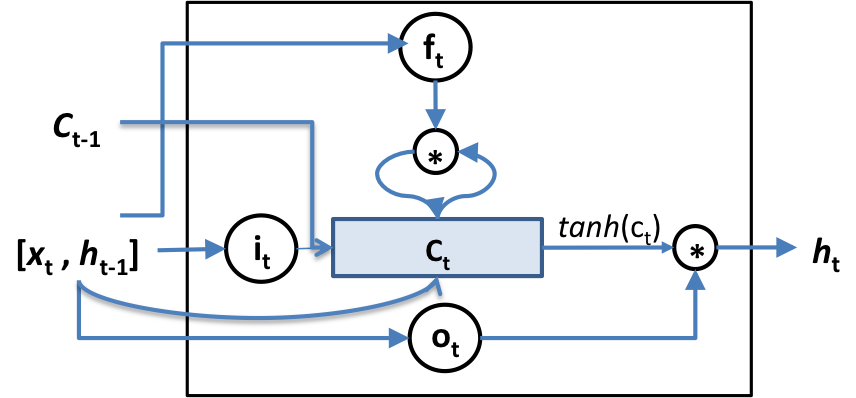}
	\end{center}
	\caption{LSTM unit used in our network. The LSTM receives as input a character $x_t$ and the value of the previous state $h_{t-1}$, and from operations in the input gate $i_t$, the forget gate $f_t$ and the output gate $o_t$ produce the next state $h_t$.}
	\label{fig:lstm}
\end{figure}

The recurrent unit used in our proposed network is the long short-term memory (LSTM) unit \cite{hochreiter@nn1997}. It addresses the main shortcomings of the original recurrent units, namely vanishing gradients (for small weights) and exploding gradients (for large weights). LSTMs are composed of a memory cell and gates that regulate the interaction of the memory cell with the input (input gate), the output (output gate) and itself (forget gate). The unit used in our network is shown in Figure~\ref{fig:lstm}. Given the input $x_t$ and the vector of the previous step $h_{t-1}$, the unit computes the vector of the current step $h_t$ based on the gates and the memory cell. The formulas to compute the values of input gate ($i_t$), forget gate ($f_t$), output gate ($o_t$), the memory cell ($c_t$) and $h_t$ are:

\begin{equation}
i_t=\sigma(W_{xi} \cdot x_t + W_{hi} \cdot h_{t-1} +  b_i)
\end{equation}

\begin{equation}
f_t=\sigma(W_{xf} \cdot x_t + W_{hf} \cdot h_{t-1} +  b_f)
\end{equation}

\begin{equation}
\tilde{c}_t= \tanh(W_{xc} \cdot x_t + W_{hc} \cdot h_{t-1} + b_c)
\end{equation}

\begin{equation}
c_t= f_t \ast c_{t-1} + i_t \ast \tilde{c}_t
\end{equation}

\begin{equation}
o_t=\sigma(W_{xo} \cdot x_t + W_{ho} \cdot h_{t-1} + b_o)
\end{equation}

\begin{equation}
h_t=o_t \ast \tanh(c_t)
\end{equation}

$W_{xi}$,$W_{xf}$,$W_{xc}$ and $W_{xo}$ are weight matrices; $b_i$,$b_f$,$b_c$ and $b_o$ are bias vectors; and $\ast$ is an element-wise multiplication. The unit we used in our network is a variation of the original LSTM. 
In the original one, the output gate also depends on $c_t$.
This modification allows the computation of $i_t$, $f_t$, $\tilde{c}_t$ and $o_t$ to be executed in parallel.

\vspace{.8cm}
\subsection{Hidden Layer with Dropout}
The last layer in both branches is the hidden layer with dropout~\cite{srivastava@jmlr2014}. Dropout is used mainly for regularization. A proportion of units in a dense layer is randomly removed in each pass of each training example with a probability $p$. By doing so, different architectures of the network with shared weights are trained. In the feed-forward propagation of a regular hidden layer, given an input vector $z$, the output is calculated as:
\begin{equation}
y = W \cdot z + b
\end{equation}
where $W$ is a weight matrix and $b$ a bias value.

Using dropout, the feed-forward propagation becomes:  
\begin{equation}
y = W \cdot (r \ast z) + b
\end{equation}
where $r$ is a vector of independent Bernoulli random variables in which each element has probability p of being 1, and $\ast$ is a element-wise multiplication. Backpropagation is only performed on the units with $p>0$. At test time, 
the weight matrix $W$ is scaled by p:
\begin{equation}
W_{test} = pW 
\end{equation}

$W_{test}$ is used at test time to score unseen examples. The value of $p$ is a user-specified hyper-parameter.

\subsection{Pooling Layer}
Previous approaches have combined hybrid networks with pooling for computer vision and video recognition tasks~\cite{lin@iccv2015,simonyan@nips2014}. Similar to them, we use a pooling layer to combine the CNN and RNN branches for our specific task of attribute annotation. 
The CNN and RNN branches execute a series of operations over the attribute value generating two different representations of it. 
A simple approach for combining them would be to concatenate the two vectors, passing the concatenated vector to the last layer, the softmax layer. Instead, we use pooling operations to combine the two branches, which try to capture their most useful patterns. 

Given two vectors $U$ and $V$, the pooling operations are:
\begin{itemize}
\item $max(U,V)$: element-wise max of $U$ and $V$. 
\item $sum(U,V)$: element-wise sum of $U$ and $V$.
\item $avg(U,V)$: element-wise average of $U$ and $V$.
\item $mul(U,V)$:  element-wise product of $U$ and $V$. 
\item $outer(U,V)$: outer product of $U$ and $V$. 
\end{itemize}

The size of $U$ and $V$ for the $max$, $sum$, $avg$ and $mul$ operations must be the same, and the size of their output vector is equal to the size of the input vectors. With respect to the outer operation, the output matrix $||U||  \times ||V||$ is flatten to a vector of size $||U||*||V||$.
This is necessary because the next layer of the network, the softmax layer, requires a 1-row vector.

\subsection{Training}
The output of the network is the class-membership probabilities provided by the softmax of a given input $x$:
\begin{equation}
P(Y=i|x,W,b)=\frac{e^{W_ix+b}}{\sum_j e^{W_jx+b}} 
\end{equation}
where $i$ is the index of a class, $W$ is a matrix of weights and $b$ is a bias term. 
The optimal parameters of the network are learned by minimizing a loss function. In our network, we use the negative log-likelihood loss function:
\begin{equation}
\mathcal{L}(X)=-\sum_{x \in X} log(P(Y=y^*|x))
\end{equation}
where $X$ is the set of training examples $x$ and $P(Y=y^*|x)$ is the probability of $x$ belonging to the true class $y^*$ returned by the softmax function. This is a commonly used loss function for multi-class classification problems. We use the stochastic optimization method Adam~\cite{kingma@iclr2015} to minimize the loss function with a learning rate of $10^{-6}$.

\section{Experimental Evaluation}\label{sec:evaluation}
In this section, we evaluate our deep learning strategy for attribute annotation, comparing it with baselines.

\begin{table}[t]
\centering
\begin{tabular}{|c|c|c|c|}  
 \hline
 \textbf{Domain} & \textbf{No. of} & \textbf{No. of} & \textbf{No. of}\\  
        & \textbf{sources} & \textbf{attributes} & \textbf{records} \\  
 \hline\hline
Car & 5 & 14 & 38,473 \\ 
Weather & 4 & 12 & 4,642 \\
Phone & 3 & 11 & 40,000 \\
Citation & 2 & 8 & 2,327 \\ 
  \hline
\end{tabular}
\caption{Description of the datasets used in our evaluation.}
\label{tab:data}
\end{table}

\subsection{Experimental Setup}
\myparagraph{Data.} We perform the evaluation on four different domains:
\begin{itemize}
\item Car: we collected structured information about cars from 5 websites\footnote{The websites for the car domain are: \url{http://www.cars.com}, \url{http://www.carfax.com}, \url{http://www.usedcars.com}, \url{http://www.truecar.com} and \url{http://www.carsforsale.com}.}. 
To gather the data, we submitted queries to their online databases. All these sites provide forms that allow queries based on zip codes. To obtain a diverse set,
we used a random list of zip codes obtained from RandomLists site\footnote{\url{https://www.randomlists.com/}} to issue queries. From the resulting pages, we extracted the attributes for all the car listings. 
Examples of attributes in this domain are: price, mileage and exterior color.
\item Weather: this dataset was created from four weather Web sites by Ambit et al.~\cite{ambite@iswc2009}. It contains attributes such as date, location and temperature. 
\item Phone: the phone dataset was obtained from online phone directories. It contains three sources and was also created by Ambit et al.~\cite{ambite@iswc2009}. Person name, address and phone are examples of attributes in this dataset.
\item Citation: the citation dataset was composed of two sources: the Cora Information Extraction dataset\footnote{\url{https://people.cs.umass.edu/~mccallum/data.html}} and the PersonalBib dataset\footnote{\url{https://www.cse.iitb.ac.in/~sunita/data/}}. Both datasets contain labeled segments for authors, title, year etc. 
\end{itemize}

Table~\ref{tab:data} presents an overview of the datasets.  The number of attributes range from 8 in the citation domain to 14 in the car domain. The domains also have a wide variation in terms of the number of records. The citation domain has 2,327 records whereas the phone domain has 40,000 records.

\begin{table*}[t]
\centering
\begin{tabular}{l | l} 
 \textbf{Hyper-Parameter} &  \textbf{Value}\\  
 \hline
Activation function & Relu \\ 
Dropout probability & 0.25\\ 
Learning rate & $10^{-6}$\\ 
Loss function & negative log-likelihood \\ 
Optimization method & Adam\\ 
Minibatch & 10\\
Epochs & 20\\
\end{tabular}
\caption{Values of the hyper-parameters used by the DL networks.}
\label{tab:parameters}
\end{table*}

\begin{table*}[t]
\centering
\begin{tabular}{c c|c|c|c|c} 
   & & \textbf{Car} & \textbf{Weather} & \textbf{Phone} & \textbf{Citation} \\  
 \hline
& MLP & 0.665 & 0.774 & 0.861 & 0.604 \\ 
\textbf{Baselines} & ONDUX & 0.708  & 0.632 & 0.852 & 0.616\\
& CRF & 0.81 & 0.886 & 0.708 & 0.812\\
\hline
& LSTM & 0.818 [e=200] & 0.872 [e=100]  & 0.873 [e=200] & 0.703 [e=100] \\ 
 & CNN & 0.855 [w=5,e=100]  & 0.903 [w=5,e=200]  & \textbf{0.876} [w=3,e=100] & 0.82 [w=5,e=100]\\ 
& concat(CNN,LSTM)  & 0.842 [w=3,e=100] & 0.913 [w=5,e=200] & 0.875 [w=3,e=100] & 0.829 [w=5,e=100]\\ 
\textbf{DL} & max(CNN,LSTM) & \textbf{0.862} [w=3,e=100] & 0.923 [w=5,e=200] & \textbf{0.876} [w=3,e=100] & 0.834 [w=5,e=300]\\ 
\textbf{Approaches} & sum(CNN,LSTM) & 0.851 [w=3,e=300] & 0.902 [w=5,e=300] & 0.875 [w=5,e=100] & 0.838 [w=5,e=100]\\ 
 & avg(CNN,LSTM) & 0.854 [w=3,e=100] & \textbf{0.924}  [w=5,e=200]  & \textbf{0.876} [w=5,e=100] & 0.84 [w=5,e=100]\\ 
& mul(CNN,LSTM) & 0.859 [w=3,e=300] & 0.902  [w=3,e=200]  & 0.875 [w=3,e=100] & \textbf{0.865} [w=5,e=100]\\ 
& outer(CNN,LSTM) & 0.855 [w=5,e=100] & 0.901[w=5,e=300] & \textbf{0.876} [w=3,e=100] & 0.823 [w=5,e=100]\\ 
 \hline
\end{tabular}
\caption{Accuracy of all approaches for the four domains. For each DL approach, we show between brackets the best accuracy and the values of window $w$ and embedding size $e$  that led to this result.}
\label{tab:results}
\end{table*}

\myparagraph{Evaluation Setup.}
For evaluation purposes, we have manually created a mapping between the attributes and the sources. We use a leave-one-out approach to evaluate the approaches: $n-1$ sources are used for training and the remaining one for testing. For each domain, we executed $n$ runs, and in each run a different source is chosen as the test source. At the end of the $n$ runs, we calculate the average accuracy. According to our definitions in Section~\ref{sec:problem}, the training sources compose the Domain Catalog for that specific run and the test source provides entities for the attribute annotation given their attributes' values. 
It is important to point out that there might be cases in which the attributes in the test source are not present in the training sources, which is something problematic for supervised learning techniques as we show later in this section.

\myparagraph{Strategies.} We implemented the following strategies for the task of attribute detection:
\begin{itemize}
\item MLP: we executed a multilayer perceptron with 2 hidden layers (300 and 50 units respectively) and a softmax layer on top. The features are bag of tokens, i.e., no consideration about the order of the characters is made. We ran 20 epochs and a minibatch size of 10.

\item ONDUX (matching): ONDUX is an approach that was originally proposed to perform attribute extraction from unstructured text. It comprises 3 steps: blocking, which performs the text tokenization; matching, in which the tokenized blocks are matched against a Domain Catalog, and reinforcement, 
which takes into consideration the order of the blocks to predict their labels. Since we are dealing with structured data whereby the tokens of each field are separated in the text and the order of attributes is in many cases arbitrary, we only implemented the matching step for comparison. ONDUX's matching step uses different functions to match a token or set of tokens to a type in the Domain Catalog. The textual matching uses a probabilistic version of $tdidf$ to match attributes' values of only textual attributes. The numerical matching matches attributes' values containing only numbers. For that, it calculates the average and standard deviation of the numerical attribute, and then evaluates how close a given value is to the distribution of the numerical attribute using a Gaussian kernel. Since many times numerical types contain non-numerical characters as, for instance, phone numbers, we normalize them, removing those characters. We do not implement the URL and email functions since these types do not appear in our datasets.

\item CRF: we implemented the CRF approach proposed by Goel at al.~\cite{goel2012exploiting} for attribute annotation. Conditional Random Fields~\cite{lafferty2001conditional} use probabilistic models for segmentation and labeling of sequences of data. It combines the strength of Hidden Markov Models (HMM) and maximum entropy Markov Models (MEMM). It models the conditional probability distribution of labels by their values, both represented as random variables. To perform the annotation, their approach: (1) splits the attribute value in tokens; (2) uses an ontology of sub-types to define the CRF structure; and (3) extracts features from those tokens to calculate a belief for each field to correspond to a certain label. There are three types of features for a token: alphabetic features (e.g starting letter and the first letter is uppercase), numeric features (e.g. starting digit and negative number) and symbol features (i.e. the symbol itself). In our evaluation, we use the approach that does not rely on an ontology of sub-types (step 2)
since we do not have access to any existing pre-built ontologies for our domains. 

\item DL approaches: these are the deep learning techniques presented in Section~\ref{sec:network}. The activation function used in the hidden layer is the rectifier function $f(x)=Max(0,x)$~\cite{nair@icml2010}. The rectifier function is more efficient than more conventional activation functions (e.g. sigmoid and hyperbolic tangent) without much difference in accuracy.
For the dropout, the probability of a node being removed is 25\%.  
In addition to pooling, we also ran the CNN and LSTM branches separately by adding a softmax layer on top of each of them and using the same parameters as the ones used for pooling. We tried 3 different embedding sizes --  100, 200 and 300 -- and 2 values of window size for the CNN branch: 3 and 5. For each network configuration, we  executed 20 epochs with a minibatch size of 10. The model with the best performance on the validation set (20\% of the training data) was the one chosen for evaluation. Table~\ref{tab:parameters} summarizes the values of hyper-parameters used in our experiments.
We used Keras~\cite{chollet2015keras} to implement our networks.
\end{itemize}

As we intend to use only the raw text to perform the classification, we did not do any pre-processing over the datasets' tokens for the deep learning approaches. For ONDUX, we normalized numerical values, as mentioned before.

\subsection{Assessing the DL Approches}
Table~\ref{tab:results} presents the accuracy obtained with each approach on the four  domains. For the CNN-based approaches (CNN and hybrid), we present the values of window $w$ and embedding size $e$ that led to the best results and for LSTM only the embedding size $e$.
As the results show, our deep learning approach obtained the best performance for all four domains. There was not a single pooling strategy, though, that outperformed the other ones: max pooling had the best accuracy for the car domain (0.862); average pooling for the weather domain (0.924); and multiplication operator for the citation domain (0.865). Regarding the phone domain, all the DL approaches obtained very close results, from 0.873 to 0.876.
Comparing the performance of the CNN versus LSTM, CNN clearly had better results in all domains. But, as we mentioned before, the combination of CNN with LSTM outperformed them in isolation.
Another interesting result is that the most suitable values of window and embedding size vary across domains but seem to be consistent for the same domain across the DL approaches. A window size of  3 and embedding size of 100 obtained the best results in the car domain, whereas in the wheather domain the best results were with window size equals to 5 and embedding size equals to 200.

In comparison to the baselines, almost all our deep learning approaches had better results than the baselines. The exceptions were: CRF obtained higher accuracy than LSTM in the weather domain (0.886 vs 0.872) and also in the citation domain (0.812 vs 0.703).  But overall, our strategy was consistently better than the baselines.  It is important to point out that our deep learning strategies do not use any hand-crafted features as CRF, neither uses different strategies for textual and numerical data as does ONDUX. By relying solely on the raw text of the attribute's value to make predictions, they obtained superior performance.

\begin{table}[t]
\centering
\begin{tabular}{|l|c|c|c|} 
 \hline
\textbf{Attribute} & \textbf{Precision} & \textbf{Recall} & \textbf{F-Measure} \\
\hline
Exterior Color & 0.792 & 0.653 & 0.716 \\
\hline
Interior Color &	0.571 & 0.877 & 0.692 \\
\hline
MPG  & 0.812 & 0.794 & 0.803 \\
\hline
Price &	1 & 0.985 & 0.992 \\
\hline
Body Style &	0.977 & 0.898 & 0.936 \\
\hline
Mileage &	0.848	& 0.601 &	0.704 \\
\hline
Transmission &	0.987 & 0.964 & 0.975 \\
\hline
VIN &	0.994 & 1 & 0.997 \\
\hline
Drive Type &	0.983 & 0.996 & 0.99 \\
\hline
Stock number &	0.657	& 0.897	& 0.758 \\
\hline
Engine & 0.849 & 0.987 & 0.913 \\
\hline
Fuel & 0.957 & 0.759 & 0.847 \\
\hline
Doors & 0.99 & 0.992 & 0.991 \\
\hline
Location & 0 & 0 & 0 \\
\hline
\end{tabular}
\caption{Precision, recall of F-Measure of the attributes in the car domain obtained by the network max(CNN,LSTM)[w=3,e=100].}
\label{tab:res_car}
\end{table}

To better understand the strengths and weaknesses of our approach, we present in Table~\ref{tab:res_car} the precision, recall and F-Measure for each attribute in the car domain obtained by the best approach in this domain: max(CNN,LSTM) [w=3,e=100]. The first thing to note is that for the location attribute all the values are 0. The reason for that is that this attribute only occurs in one out of the five sources ($truecar.com$) of the car domain. Therefore, in the run in which the $truecar.com$ is the test source, the location attribute does not appear in the training data (or Domain Catalog), since none of the remaining four sources has this attribute. As a result of that, not a single value of the location attribute was correctly classified.
There are also cases whereby the attributes have overlap of values, which hurts the  performance of the classifier. The attributes exterior color and interior color are examples of that with F-Measure values of 0.716 and 0.692 respectively. Values such as ``black'', ``white'' and ``gray'' are common in both.
Since our sequence-based network relies on the format of the values of the attributes to make the prediction, it is an issue when the format of an attribute in the training sources differs from the test source. For instance, the format of the mileage attribute in the website $carsforsale$ is completely different from the other sources, which resulted in an F-Measure of 0.706. Usually in the car sources, this attribute is represented as, e.g., ``21/32'' whereas in $carsforsale$ its format is something like  ``14 city / 19 hwy EPA Fuel Economy Guide''. 
As expected, our approach obtained good results for attributes with similar format and distribution across sources regardless whether they are textual (e.g. transmission) or numerical (e.g. price and the vehicle identification number - VIN). 
The main cause of all these issues is the mismatch between the training data (or Domain Catalog) and the test sources. Having the Domain Catalog a representative sample of the instances in the domain, this mismatch can be mitigated.

\section{Related Work}\label{sec:related}
This work is the first to propose a hybrid deep learning network for the problem of attribute annotation. In this section, we discuss previous approaches for the attribute annotation problem and hybrid deep learning networks that have been proposed for other problems in text and image processing.
\subsection{Attribute Annotation}
%
ONDUX \cite{cortez2010ondux} is a three-step based approach, as briefly explained in section \ref{sec:evaluation}.  The steps involved in this approach are called Blocking, Matching and Reinforcement. Each one of them refines the results achieved by the previous step. Blocking segments the text
based on the co-occurrence of a term in the input and in the Knowledge Base\footnote{A Knowledge Base is a set  $K = \{\langle a_1, O_1\rangle, \ldots, \langle a_n, O_n\rangle\}$, where each $a_i$ is an attribute and each $O_i$ is a set of strings}. The Matching step consists of associating the block from the previous step to the attributes in the Knowledge Base. This step handles textual and numerical fields in different ways. For textual matching, each block is compared to the occurrences in the Knowledge Base through a similarity function similar to \emph{tfidf}~\cite{salton1986introduction}. If the block is numeric, it treats the numerical occurrences as samples from a Gaussian Distribution and estimates how close the values in the block are from the mean of that distribution. The last step, Reinforcement, uses a probabilistic graphical model much alike HMM~\cite{rabiner1986introduction} to correct mislabeled fields from the matching step. As we stated in our problem setting, since we are dealing with structured data whereby the tokens of each field are separated in the text and the order of attributes is in many cases arbitrary, only the Matching step applies to the problem we are trying to solve. 

Limaye et al.~\cite{limaye2010annotating} proposed an approach based on probabilistic graphical models to annotate simultaneously entities, types and relations in tables. To perform this task, they map types in a given ontology to values in a column of a table.
The model is based on Factor Graph \cite{koller2009probabilistic}. The two main design decisions of this model consist on the choice of random variables and its values, and the construction of potential functions, a.k.a factors. There are random variables for the type of each column in the data, for the label of each cell and for each relation between two columns. The chosen factors are a function of the weighted sum of features. A major drawback of this approach is its requirement of having an ontology in the domain, which is not always possible or practical.

Goel et al.~\cite{goel2012exploiting} proposed a CRF-based approach for the problem of attribute annotation. For that, it exploits the structure within the data in each attribute value. 
In their context, each cell is called a field and each field is split into tokens. They propose three arrangements of variables: the first one only takes in consideration the field features; the second one takes in consideration the field features and also the token features; and the third one takes in consideration the previous features and captures the idea of order between tokens.
If one takes in consideration the order among fields, then we can replicate the previous three kind of graphs and end up with six different graphs that exploit increasing levels of complexity of the data. 
To implement the second and third arrangements, an ontology of sub-types which is rarely available is required.

\subsection{Hybrid Deep Learning}
Santos et al.~\cite{santos@acl2015} proposes a hybrid deep learning representation for the problem of equivalent questions. More specifically, their network models each question with a CNN and a bag-of-words representation. They combine the hybrid representations using a linear combination of the cosine similarity of the pair of questions in each type of network. 
Hybrid representations have also been used in the areas of computer vision~\cite{lin@iccv2015} and video recognition~\cite{parkcombining,simonyan2014two}.
Lin at al. \cite{lin@iccv2015} use billinear CNN for visual recognition. A billinear model consists of two feature extractors, in this case CNNs, and their outputs are combined through outer product to obtain an image descriptor. 
Park et al.~\cite{parkcombining} combines multiple CNNs trained in different sources for the task of action recognition. They used two CNNs branch: one to model spatial information and the other to model temporal information in videos. The branches are combined using the element-wise product. 
We use a similar approach by pooling two different branches (an RNN branch and a CNN branch) to capture the sequence structure of attributes' values.

\section{Conclusions}\label{sec:conc}
In this paper, we present a hybrid deep learning strategy for the problem of annotating attributes from web entities based only on their values. For that, it captures the sequence structure of the characters that compose the attributes' values without relying on any hand-crafted features nor pre-processing of the input. Our network combines two types of sequence-based neural networks namely, a convolutional neural network and a recurrent neural network, that produce different representations of the same input. These representations are then combined in a pooling layer that captures their most important features. We have shown in the experimental evaluation that: (1) our DL strategy outperforms the existing approaches for this task; and (2) the CNN branch has a better performance than the RNN branch.
Since our hybrid network is very modular, possible extensions to this work would be to apply other types of deep learning units in our network.

\hidecomment{
\begin{itemize}
\item Comparison between deep learning and CRF
\item Restrictions and suppositions
\item Training is expensive but robustness pays off
\item Future work
\begin{itemize}
    \item New NN structures to test
    \item New data sets (and the restrictions in the current ones)
    \end{itemize}
\end{itemize}
}
\bibliographystyle{abbrv}
\bibliography{paper}

\begin{thebibliography}{10}

\bibitem{ambite@iswc2009}
J.~L. Ambite, S.~Darbha, A.~Goel, C.~A. Knoblock, K.~Lerman, R.~Parundekar, and
  T.~Russ.
\newblock Automatically constructing semantic web services from online sources.
\newblock In {\em Proceedings of the 8th International Semantic Web
  Conference}. Springer, 2009.

\bibitem{cafarella@sigrecord2008}
M.~Cafarella, E.~Chang, A.~Fikes, A.~Halevy, W.~Hsieh, A.~Lerner, J.~Madhavan,
  and S.~Muthukrishnan.
\newblock Data management projects at google.
\newblock {\em ACM SIGMOD Record}, 37(1):34--38, 2008.

\bibitem{cafarella@cacm2011}
M.~J. Cafarella, A.~Halevy, and J.~Madhavan.
\newblock Structured data on the web.
\newblock {\em Communications of the ACM}, 54(2):72--79, 2011.

\bibitem{cafarella@vldb2008}
M.~J. Cafarella, A.~Halevy, D.~Z. Wang, E.~Wu, and Y.~Zhang.
\newblock Webtables: exploring the power of tables on the web.
\newblock {\em Proceedings of the VLDB Endowment}, 1(1):538--549, 2008.

\bibitem{chollet2015keras}
F.~Chollet.
\newblock Keras: Theano-based deep learning library.
\newblock {\em Code: https://github. com/fchollet. Documentation: http://keras.
  io}, 2015.

\bibitem{collobert@2011}
R.~Collobert, J.~Weston, L.~Bottou, M.~Karlen, K.~Kavukcuoglu, and P.~Kuksa.
\newblock Natural language processing (almost) from scratch.
\newblock {\em The Journal of Machine Learning Research}, 12:2493--2537, 2011.

\bibitem{cortez2010ondux}
E.~Cortez, A.~S. da~Silva, M.~A. Gon{\c{c}}alves, and E.~S. de~Moura.
\newblock Ondux: on-demand unsupervised learning for information extraction.
\newblock In {\em Proceedings of the 2010 ACM SIGMOD International Conference
  on Management of data}, pages 807--818. ACM, 2010.

\bibitem{crestan@wsm2011}
E.~Crestan and P.~Pantel.
\newblock Web-scale table census and classification.
\newblock In {\em Proceedings of the fourth ACM international conference on Web
  search and data mining}, pages 545--554. ACM, 2011.

\bibitem{cruse2006neural}
H.~Cruse.
\newblock Neural networks as cybernetic systems.
\newblock {\em Brain, minds, and media. See http://www. brains-minds-media.
  org/archive/289}, 2006.

\bibitem{santos@acl2015}
C.~dos Santos, L.~Barbosa, D.~Bogdanova, and B.~Zadrozny.
\newblock Learning hybrid representations to retrieve semantically equivalent
  questions.
\newblock In {\em Proceedings of the 53rd Annual Meeting of the Association for
  Computational Linguistics}, pages 694--699, 2015.

\bibitem{goel2012exploiting}
A.~Goel, C.~A. Knoblock, and K.~Lerman.
\newblock Exploiting structure within data for accurate labeling using
  conditional random fields.
\newblock In {\em Proceedings on the International Conference on Artificial
  Intelligence (ICAI)}, page~1. The Steering Committee of The World Congress in
  Computer Science, Computer Engineering and Applied Computing (WorldComp),
  2012.

\bibitem{gupta@vldb2009}
R.~Gupta and S.~Sarawagi.
\newblock Answering table augmentation queries from unstructured lists on the
  web.
\newblock {\em Proceedings of the VLDB Endowment}, 2(1):289--300, 2009.

\bibitem{hochreiter@nn1997}
S.~Hochreiter and J.~Schmidhuber.
\newblock Long short-term memory.
\newblock {\em Neural computation}, 9(8):1735--1780, 1997.

\bibitem{kim1@emnlp2014}
Y.~Kim.
\newblock Convolutional neural networks for sentence classification.
\newblock In {\em Proceedings of Conference on Empirical Methods in Natural
  Language Processing (EMNLP)}, 2014.

\bibitem{kingma@iclr2015}
D.~P. Kingma and J.~B. Adam.
\newblock A method for stochastic optimization.
\newblock In {\em International Conference on Learning Representation}, 2015.

\bibitem{koller2009probabilistic}
D.~Koller and N.~Friedman.
\newblock {\em Probabilistic graphical models: principles and techniques}.
\newblock 2009.

\bibitem{lafferty2001conditional}
J.~Lafferty, A.~McCallum, and F.~C. Pereira.
\newblock Conditional random fields: Probabilistic models for segmenting and
  labeling sequence data.
\newblock 2001.

\bibitem{lecun@ieee1998}
Y.~LeCun, L.~Bottou, Y.~Bengio, and P.~Haffner.
\newblock Gradient-based learning applied to document recognition.
\newblock {\em Proceedings of the IEEE}, 86(11):2278--2324, 1998.

\bibitem{limaye2010annotating}
G.~Limaye, S.~Sarawagi, and S.~Chakrabarti.
\newblock Annotating and searching web tables using entities, types and
  relationships.
\newblock {\em Proceedings of the VLDB Endowment}, 3(1-2):1338--1347, 2010.

\bibitem{lin@iccv2015}
T.-Y. Lin, A.~RoyChowdhury, and S.~Maji.
\newblock Bilinear cnn models for fine-grained visual recognition.
\newblock In {\em Proceedings of the IEEE International Conference on Computer
  Vision}, pages 1449--1457, 2015.

\bibitem{madhavan@cidr2007}
J.~Madhavan, S.~Jeffery, S.~Cohen, X.~Dong, D.~Ko, C.~Yu, and A.~Halevy.
\newblock Web-scale data integration: You can only afford to pay as you go.
\newblock In {\em Proceedings of CIDR}, pages 342--350, 2007.

\bibitem{meusel@cikm2014}
R.~Meusel, P.~Mika, and R.~Blanco.
\newblock Focused crawling for structured data.
\newblock In {\em Proceedings of the 23rd ACM International Conference on
  Conference on Information and Knowledge Management}, pages 1039--1048. ACM,
  2014.

\bibitem{nair@icml2010}
V.~Nair and G.~E. Hinton.
\newblock Rectified linear units improve restricted boltzmann machines.
\newblock In {\em Proceedings of the 27th International Conference on Machine
  Learning (ICML-10)}, pages 807--814, 2010.

\bibitem{parkcombining}
E.~Park, X.~Han, T.~L. Berg, and A.~C. Berg.
\newblock Combining multiple sources of knowledge in deep cnns for action
  recognition.
\newblock In {\em Winter Conference on Applications of Computer Vision (WACV)},
  2016.

\bibitem{qiu@vldb2015}
D.~Qiu, L.~Barbosa, X.~L. Dong, Y.~Shen, and D.~Srivastava.
\newblock Dexter: Large-scale discovery and extraction of product
  specifications on the web.
\newblock {\em Proceedings of the VLDB Endowment}, 8(13):2194--2205, 2015.

\bibitem{rabiner1986introduction}
L.~R. Rabiner and B.-H. Juang.
\newblock An introduction to hidden markov models.
\newblock {\em ASSP Magazine, IEEE}, 3(1):4--16, 1986.

\bibitem{salton1986introduction}
G.~Salton and M.~J. McGill.
\newblock Introduction to modern information retrieval.
\newblock 1986.

\bibitem{simonyan@nips2014}
K.~Simonyan and A.~Zisserman.
\newblock Two-stream convolutional networks for action recognition in videos.
\newblock In {\em Advances in Neural Information Processing Systems}, pages
  568--576, 2014.

\bibitem{simonyan2014two}
K.~Simonyan and A.~Zisserman.
\newblock Two-stream convolutional networks for action recognition in videos.
\newblock In {\em Advances in Neural Information Processing Systems}, pages
  568--576, 2014.

\bibitem{srivastava@jmlr2014}
N.~Srivastava, G.~Hinton, A.~Krizhevsky, I.~Sutskever, and R.~Salakhutdinov.
\newblock Dropout: A simple way to prevent neural networks from overfitting.
\newblock {\em The Journal of Machine Learning Research}, 15(1):1929--1958,
  2014.

\bibitem{sutskever@nips2014}
I.~Sutskever, O.~Vinyals, and Q.~V. Le.
\newblock Sequence to sequence learning with neural networks.
\newblock In {\em Advances in neural information processing systems}, pages
  3104--3112, 2014.

\end{thebibliography}

\end{document}